%% file: TopProceedings.tex
\documentclass[12pt]{article}
\usepackage{graphicx}
\usepackage{amsmath, amssymb}
\usepackage{cite}

\newcommand\pubnumber{}
\newcommand\pubdate{\today}

\def\institute{Institute for Particle Physics and Astrophysics\\
Eidgen\"ossische Technische Hochschule Z\"urich, 8093 Z\"urich, SWITZERLAND}

\def\Title#1{\begin{center} {\Large #1 } \end{center}}
\def\Author#1{\begin{center}{ \sc #1} \end{center}}
\def\Address#1{\begin{center}{ \it #1} \end{center}}

\newcommand\pubblock{\rightline{\begin{tabular}{l} \pubnumber\\
         \pubdate  \end{tabular}}}
\newenvironment{Abstract}{\begin{quotation}  }{\end{quotation}}
\newenvironment{Presented}{\begin{quotation} \begin{center} 
             PRESENTED AT\end{center}\bigskip 
      \begin{center}\begin{large}}{\end{large}\end{center} \end{quotation}}
\def\Acknowledgements{\bigskip  \bigskip \begin{center} \begin{large}
             \bf ACKNOWLEDGEMENTS \end{large}\end{center}}

\input econfmacros.tex

\begin{document}
\begin{titlepage}
\pubblock

\def\pt{p$_T$}
\def\tthbb{t$\bar{\text{t}}$H, H$\rightarrow$b$\bar{\text{b}}$}
\def\ttbb{t$\bar{\text{t}}$+jets}
\def\tt2b{t$\bar{\text{t}}$+bb}
\def\tth{t$\bar{\text{t}}$H}

\vfill
\Title{Reconstruction of \tthbb{} events \\
using the matrix element method \\
\vspace{2mm}
and substructure techniques}
\vfill
\Author{Maren Meinhard}
\Address{\institute}
\vfill
\begin{Abstract}
This contribution outlines the implementation of the matrix element method (MEM) in the search for \tthbb{} events. In particular, the evaluation of the transfer functions, which relate detector level to parton level quantities in the computation of the MEM, is described. In addition, it is presented how jet substructure reconstruction can be combined with the MEM. The combination of these techniques leads to a decrease in computation time by up to 90\% and an increase in event selection efficiency of 30\% in the high Higgs boson and top quark \pt{} phase space.
\end{Abstract}
\vfill
\begin{Presented}
$11^\mathrm{th}$ International Workshop on Top Quark Physics\\
Bad Neuenahr, Germany, September 16--21, 2018
\end{Presented}
\vfill
\end{titlepage}
\def\thefootnote{\fnsymbol{footnote}}
\setcounter{footnote}{0}

\def\pt{p$_T$}
\def\tthbb{t$\bar{\text{t}}$H, H$\rightarrow$b$\bar{\text{b}}$}
\def\ttbb{t$\bar{\text{t}}$+jets}
\def\tt2b{t$\bar{\text{t}}$+bb}
\def\tth{t$\bar{\text{t}}$H}

\section{Introduction}

In the absence of direct observation of beyond the standard model (BSM) physics at the LHC, precision measurements are becoming the main pathway to test the consistency of the standard model (SM). In this context the associated production of a Higgs boson with a top-antitop quark pair (\tthbb) is particularly interesting, since this channel allows for a direct measurement of the top and bottom quark Yukawa couplings. Both the \tth{} production process and the Higgs boson decay to bottom quarks have been observed by the CMS and ATLAS collaborations in 2018~\cite{TTHATLAS}. The identification of \tthbb{} events is challenging due to the complicated final state involving many jets, leptons and MET, and by the irreducible background \tt2b{} whose cross section ($\sim$ 832\;pb) is over three magnitudes larger than the \tth{} cross section ($\sim$ 0.5\;pb). Furthermore the combinatorial self-background for various possibilities to associate jets to initial quarks further complicates the matter, as well as the large theoretical uncertainties on the predictions for the \tt2b{} process.

In order to overcome these difficulties, dedicated analysis techniques are necessary. In particular, the matrix element method is well suited for this analysis channel since it intrinsically solves the combinatorial background issue.

\section{The matrix element method}

The matrix element method (MEM)~\cite{MEM} is a fully analytical method which computes\footnote{The CUBA package~\cite{CUBA} is used for numerical integration.} the probability for an event with characteristics $\boldsymbol{y}$ (kinematics of jets, MET, leptons...) to originate from the underlying process $\boldsymbol{\alpha}$:

\begin{equation} \label{MEM}
P(\boldsymbol{y}|\boldsymbol{\alpha}) \propto \frac{1}{\sigma_\alpha}\int d\Phi(\boldsymbol{x})\; |M_\alpha | ^2(\boldsymbol{x}) \; W(\boldsymbol{x},\boldsymbol{y})
\end{equation}
where $\sigma_\alpha$ is the total cross section of process $\boldsymbol{\alpha}$, $d\Phi(\boldsymbol{x})$ is the phase space measure, $|M_\alpha | ^2(\boldsymbol{x})$ is the LO scattering amplitude squared and $W(\boldsymbol{x},\boldsymbol{y})$ represents the probability to obtain a detector response $\boldsymbol{y}$ for a particle level event $\boldsymbol{x}$. A discriminant can be obtained by taking the ratio of the signal $P(\boldsymbol{y}|\boldsymbol{\text{t}\bar{\text{t}}\text{H}, \text{H}\rightarrow\text{b}\bar{\text{b}}})$ and the background $P(\boldsymbol{y}|\text{t}\bar{\text{t}}+\text{bb})$ probabilities.

Since the parton level quantities $\boldsymbol{x}$ are unknown, they must be related to the observed quantities $\boldsymbol{y}$ via transfer functions $W(\boldsymbol{x},\boldsymbol{y})$. These are evaluated separately for each type of particle in the detector. Considering the much better momentum and angular resolution for leptons than for jets, a Dirac delta function models the transfer functions for leptons. For the MET, a multivariate normal distribution is used:

\begin{equation}
W_{\mathrm{MET}} (\textbf{p}_T\vert\sum_k \textbf{p}_k) = \frac{1}{2\pi \vert \Sigma \vert ^{1/2}} \exp \left[ -\frac{1}{2} (\textbf{p}_T - \sum_k \textbf{p}_k)^T  \Sigma^{-1} (\textbf{p}_T - \sum_k \textbf{p}_k) \right]
\end{equation}

with $\Sigma = \sigma_{\mathrm{MET}} \textbf{I}$ and $\sigma_{\mathrm{MET}}$ = 30 GeV.

\begin{figure}[htb]
\centering
\includegraphics[width=0.48\textwidth]{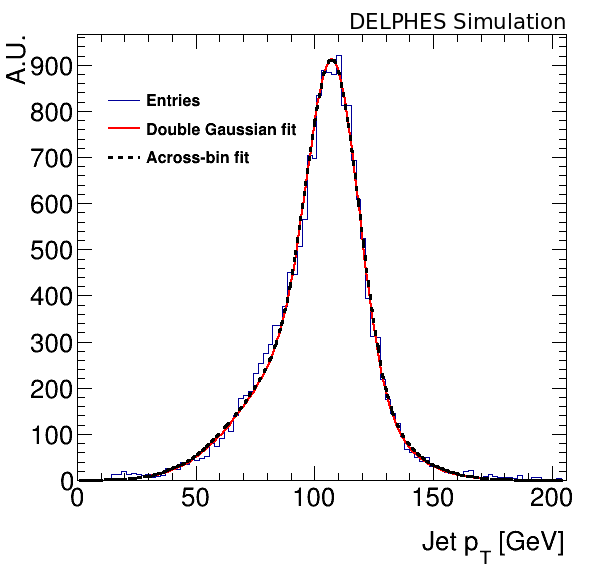} 
\includegraphics[width=0.48\textwidth]{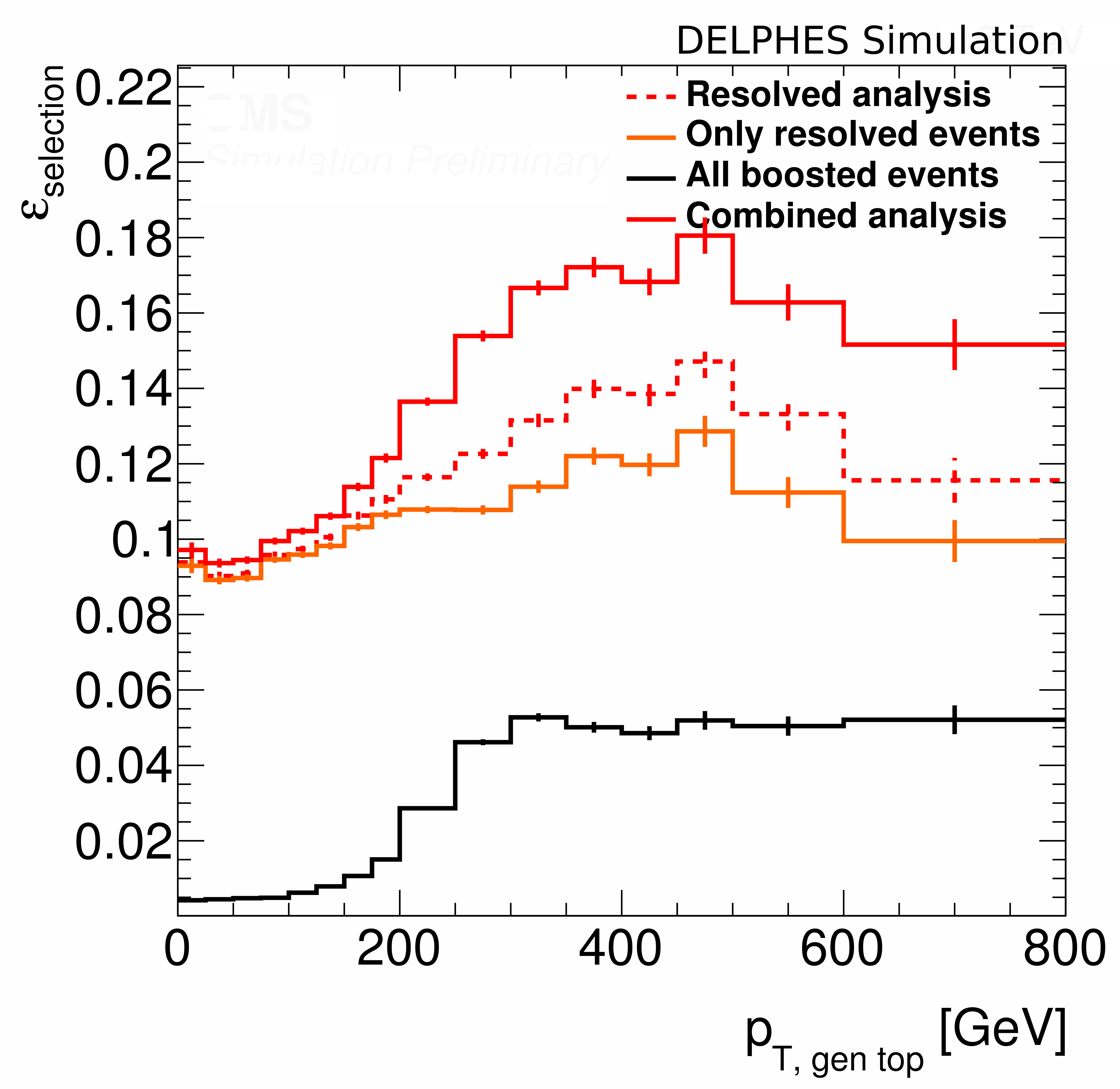} 
\caption{Double-gaussian and across-bin fits of jets originating from quarks with $p_{T,gen}$ $\in$ [100,102]\;GeV (left). Event selection efficiency as a function of generated top quark \pt{} for standard resolved analysis (red, dashed), resolved only (orange), events with HTTV2 candidate (black) and combined boosted analysis (red, solid) (right).}
\label{fig:pt} 
\end{figure}

Transfer functions for jets are evaluated from MC simulation, taking into account the varying detector resolution according to a jet's \pt, pseudorapidity and flavour. A double gaussian distribution models the jet \pt{} distribution corresponding to quarks within a specific \pt{} range ($p_{T,gen}$) as shown in Equation~\ref{TFJet} and  Figure~\ref{fig:pt} (left). An across-bin fit in \pt{} is performed for each of the four parameters of the double-gaussian fit ($\alpha_i$ for $i \in 1, \cdots ,4$). The jet transfer functions can then be reconstructed using these across-bin fits, and it can be seen in Figure~\ref{fig:pt} (left) that this reconstruction models the distribution well.

\begin{equation} \label{TFJet}
TF(p_T | p_{T,\mathrm{gen}}) = N \biggl[0.7\exp{\biggl(\frac{p_{T,gen} - p_T -\alpha_1}{\alpha_2}\biggr)^2} \\ + 0.3\exp{\biggl(\frac{p_{T,gen} - p_T -\alpha_3}{\alpha_2+\alpha_4}\biggr)^2}\biggr]
\end{equation} 

In evaluating the MEM probability, all possibilities to associate the jets to the underlying quarks need to be considered. In particular for \tthbb{} events, 12 permutations are considered to map all four b tagged jets to the b quarks produced by the decay of the Higgs boson and both top quarks. In order to perform these permutations, at least four b tagged jets are required for the MEM computation.

\section{Merging substructure techniques with the MEM}

When the top quark and/or the Higgs boson in \tthbb{} events have an increasingly high transverse momentum, their decay products become collimated and might merge into a single classical jet, therefore failing the mapping of the observed jets to the quarks in the evaluation of the MEM. This can be solved using substructure techniques. In these proceedings, emphasis will be put on tagging top quarks with high \pt. However, tagging Higgs bosons is an equally promising idea. Furthermore, only events with exactly one top quark decaying leptonically are considered.

Signal and background (\ttbb{}) events are generated using POWHEG v.2~\cite{POW1} while PYTHIA v.8.200~\cite{PYTHIA} is used to model hadronization and parton shower. The detector response is modelled with DELPHES~\cite{DELPHES}. All other specifications such as pileup contributions correspond to the 2017 LHC conditions. Jets are reconstructed with the anti-k$_T$ algorithm~\cite{ANTIKT} (R=0.4) and tagged as b quarks if they pass the loose working point (corresponding to a $\sim$10\% light-flavour misidentification rate).

The HEPTopTagger V2 algorithm (HTTV2)~\cite{HTT} is used for top quark identification. Using as input Cambridge-Aachen~\cite{CA1} large-R jets with a distance parameter of 1.5, the algorithm reconstructs exactly three subjets from the large-R jet, while removing contributions from pileup and underlying events.

\begin{figure}[htb]
\centering
\includegraphics[width=0.55\textwidth]{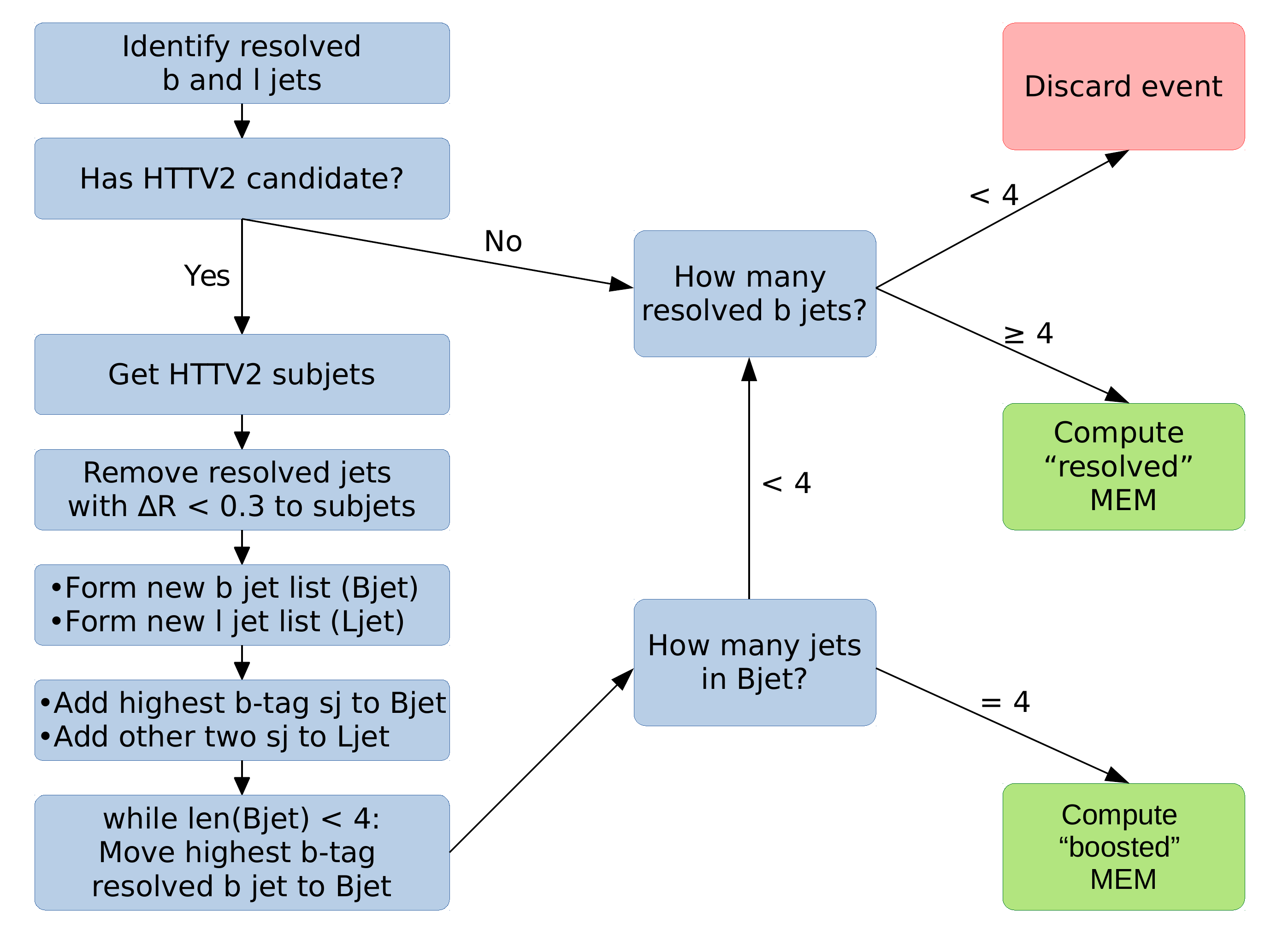}
\caption{Analysis strategy for merging top quark candidates with the MEM.}
\label{fig:boosted}
\end{figure}

Figure~\ref{fig:boosted} illustrates how information from HTTV2 candidates can be used in the MEM analysis. To do so, new input lists of jets for the MEM are created, using one b tagged subjet and two light tagged subjets from the HTTV2 candidate. Resolved b tagged jets which cannot be associated to HTTV2 subjets are added to the new input list of b jets, which is then used for the calculation of the boosted MEM. Hence resolved jets are replaced with HTTV2 subjets in the boosted MEM. For events with no HTTV2 candidate, the standard, resolved MEM can be computed. 

Combining substructure techniques with the MEM leads to several improvements. In particular the long computation time for the MEM evaluation, which is one of its major drawbacks especially when looking ahead at HL-LHC conditions, can be significantly reduced for events with a HTTV2 candidate. In the standard resolved analysis, the computation of the MEM requires 52.5 $\pm$ 27.4\;s\footnote{Evaluated using standard x86 CPU units with a Intel Xeon X5560 processor. The uncertainties represent the standard deviation.} per event. When replacing jets with HTTV2 subjets, the computation time is reduced by $\sim$~75\% to 13.7 $\pm$ 1.4\;s. This is because many events have additional jets coming from QCD radiation. Knowing which non b tagged jets arise from the top quark decay, QCD jets can be removed, thus reducing the number of permutations to consider. 

The computation time can be further reduced as the HTTV2 subjets correspond to the top quark's decay products. Therefore, only 3 permutations need to be considered to map all jets to the generator level quarks. In this case, the computation time is reduced to 3.5 $\pm$ 0.3\;s, which corresponds to a total speed-up of 90\%.

Furthermore, the newly devised analysis strategy allows to recover events for which the MEM could not be evaluated with resolved jets. This occurs when too few b tagged jets are identified, which happens when a top quark has a large \pt, and the b quark cannot be reconstructed as a single resolved jet. These events can be recovered by reconstructing the top quark's decay structure with the HTTV2 algorithm. Figure~\ref{fig:pt} (right) shows the event selection efficiency as a function of the generated top quark \pt{} for the standard resolved and the combined boosted analysis. As expected, the efficiency in the latter one exceeds the selection efficiency of the resolved analysis in the regime of top quarks with large \pt{} by up to 30\%. Since the \pt{} of the top quarks and the Higgs boson in the \tthbb{} channel are correlated, this also leads to an increased event selection efficiency for Higgs bosons with large \pt.

This is promising since many BSM theories predict deviations from the SM in the high \pt{} region of the phase space. For example, deviations from the SM CP structure of the top quark Yukawa coupling~\cite{COUPLING} could appear at large Higgs boson \pt{} or additional terms in the standard model Lagrangian~\cite{LAGRANGIAN} could lead to a modified momentum distribution of the Higgs boson. Identifying events in this phase space would therefore enable to test these theories.

\section{Conclusion}

The matrix element method is a powerful analysis technique which allows the reconstruction of complicated final states such as the \tthbb{} process. A key part of the MEM computation is the evaluation of the transfer functions. It has been shown that they are well modelled by double gaussian distributions. Furthermore, a strategy to combine substructure methods, in particular top quark identification, with the MEM has been presented. A major advantage of this method is the reduction of the computation time by up to $\sim$90\%. This comes from the reduction of the number of permutations, being able to identify QCD radiation jets, and removing part of the self-background. Furthermore, the event selection efficiency can be enhanced in the regime of top quarks and Higgs bosons with large \pt, which is crucial for testing various BSM theories which affect this region of the phase space.

\Acknowledgements
I am grateful to A. Gomez, G. Kasieczka, J. Pata and C. Rei{\ss}el for their support and constructive comments which were essential to the realization of this work.

\end{document}

%% file: econfmacros.tex
\def\beq{\begin{equation}}
\def\eeq#1{\label{#1}\end{equation}}
\def\eeqn{\end{equation}}

\def\beqa{\begin{eqnarray}}
\def\eeqa#1{\label{#1}\end{eqnarray}}
\def\eeqan{\end{eqnarray}}

\let\bar=\overbar

\def\Dslash{\not{\hbox{\kern-4pt $D$}}}
\def\dslash{\not{\hbox{\kern-2pt $\del$}}}

\def\msb{{\bar{\ssstyle M \kern -1pt S}}}

